\pdfoutput=1

\documentclass[runningheads]{llncs}
\usepackage{url}  
\usepackage{graphicx}  

\usepackage{amsmath}
\usepackage{amssymb}
\usepackage{multirow}
\usepackage{array}
\newcolumntype{L}[1]{>{\raggedright\let\newline\\\arraybackslash\hspace{0pt}}m{#1}}
\newcolumntype{C}[1]{>{\centering\let\newline\\\arraybackslash\hspace{0pt}}m{#1}}

\usepackage{pgfplots}
\usepackage{pgfplotstable}
\pgfplotsset{compat=newest}

\begin{document}
%
\title{AI Meets Austen: Towards Human-Robot Discussions of Literary Metaphor}

\author{Natalie Parde\inst{1} \and
	Rodney D.~Nielsen\inst{2}}
\authorrunning{Parde and Nielsen}
%
\institute{Department of Computer Science, University of Illinois at Chicago \and
	Department of Computer Science and Engineering, University of North Texas\\
	\email{parde@uic.edu, rodney.nielsen@unt.edu}}

\maketitle
\begin{abstract}
Artificial intelligence is revolutionizing formal education, fueled by innovations in learning assessment, content generation, and instructional delivery.  Informal, lifelong learning settings have been the subject of less attention.  We provide a proof-of-concept for an embodied book discussion companion, designed to stimulate conversations with readers about particularly creative metaphors in fiction literature.  We collect ratings from 26 participants, each of whom discuss Jane Austen's \textit{Pride and Prejudice} with the robot across one or more sessions, and find that participants rate their interactions highly.  This suggests that companion robots could be an interesting entryway for the promotion of lifelong learning and cognitive exercise in future applications.
\end{abstract}

\section{Introduction}
Robotic companions have been examined in many educational settings, acting as learning partners \cite{10.1007/978-3-319-93846-2_84}, intelligent tutors \cite{10.1007/978-3-642-39112-5_100,AAAI159280,leyzberg2012physical,Schodde:2017:ARL:2909824.3020222}, teachable agents \cite{Hood:2015:CTR:2696454.2696479,10.1007/978-3-642-39112-5_31,Tanaka:2012:CTC:3109680.3109685}, and feedback providers \cite{10.1007/978-3-319-93846-2_2}.  A common goal among most robots filling these roles to date has been the furtherance of specific learning objectives.  They have been underutilized in informal learning settings, which may call for robots to tackle fuzzier objectives for which open-ended conversation is a better avenue of interaction.  Reading is a cognitively rewarding way to engage in informal lifelong learning \cite{doi:10.1093/gerona/59.4.M390,berns2013short,mar2006bookworms,mar2009exploring,payne2012effects,Payne2014157}, but the potential for companion robots to play a role in motivating lifelong reading behaviors has remained untapped.  We set out to fill that void by developing a proof-of-concept embodied conversational companion capable of engaging readers in discussions about books.  

We select creative metaphor (a particularly cognitively demanding form of rhetoric \cite{lai2009comprehending}) as our literary focus, and demonstrate that an automatic metaphor novelty scoring approach can be harnessed to identify interesting metaphors in literature.  We design a conversational dialogue system that makes use of questions generated about those metaphors, and implement it in a companion robot.  This is the first approach, either computational or otherwise, to employ metaphor as an impetus for lifelong cognitive exercise, and the first embodied conversational system created chiefly to promote such exercise.  We empirically confirm that users of the completed prototype rate it as likeable and engaging, and maintain this sentiment over multiple sessions.  These contributions form an essential proof-of-concept for an embodied lifelong learning companion.

\section{Related Work}
Educational scenarios to which social robots have been deployed have been primarily formal settings with child learners \cite{10.1007/978-3-319-93846-2_2,10.1007/978-3-642-39112-5_100,AAAI159280,Hood:2015:CTR:2696454.2696479,10.1007/978-3-319-93846-2_84,10.1007/978-3-642-39112-5_31,Tanaka:2012:CTC:3109680.3109685}.  Our focus is on a different setting: informal, conversational lifelong learning.  \textit{Lifelong learning} is the process of acquiring knowledge and/or exercising cognitive faculties across the human lifespan, outside of traditional academic contexts. 
Research involving social robots in lifelong learning scenarios has been scarce, with most work deploying social robots to adult populations focusing on psychological or physical healthcare needs instead \cite{ElKamali:2018:TNE:3267305.3274188,4651113,piatt2017companionship,Winkle:2018:SRE:3171221.3171273}.  However, Tapus et al.~\cite{5209501} designed a human-robot music guessing game to stimulate cognition in older adults suffering from dementia, and Deublein et al.~\cite{DEUBLEIN2018182} created a social robot to scaffold motivation in adult second language learners.  Schodde et al.~\cite{Schodde:2017:ARL:2909824.3020222} also explored second language learning in adults, although their robot's behaviors were originally designed with children in mind.  A common theme across these systems is the absence of open-ended conversation: all cases utilize buttons and multiple-choice answers as their input.  The inability to converse naturally limits a robot's potential to engage in cognitively meaningful interactions, particularly when dealing with more subjective topics like literature or metaphor interpretation.

Although virtual avatars could implement the same methods as robots in most of these cases, they may fall short of achieving the same goals.  Research has demonstrated that physically embodied robots elicit longer conversations and more positive perceptions than computer agents \cite{Powers:2007:CCA:1228716.1228736}, and are better able to influence people than virtual avatars or videos of the same robots \cite{LI201523}.  In accordance with these findings, we implement our system using a physically embodied robot to maximize its anticipated utility.

\section{System Design}
\begin{figure}[t]
	\centering
	\includegraphics[width=0.9\textwidth]{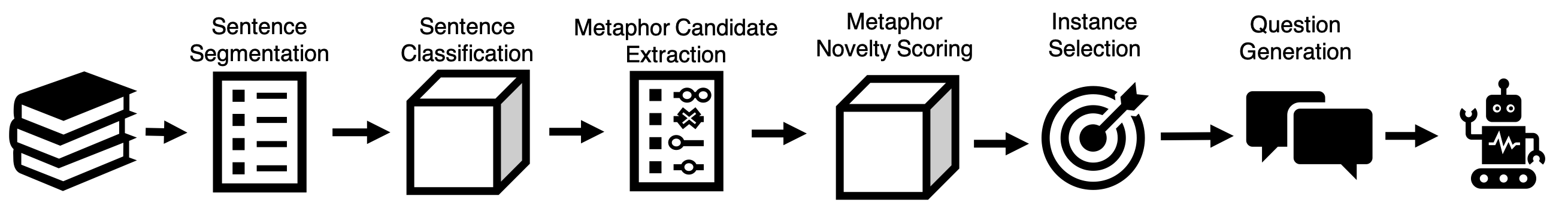}
	\caption{Processing content from raw text to a series of questions.}
	\label{processing_pipeline}
\end{figure}

\begin{figure}[t]
	\centering
	\includegraphics[width=0.7\textwidth]{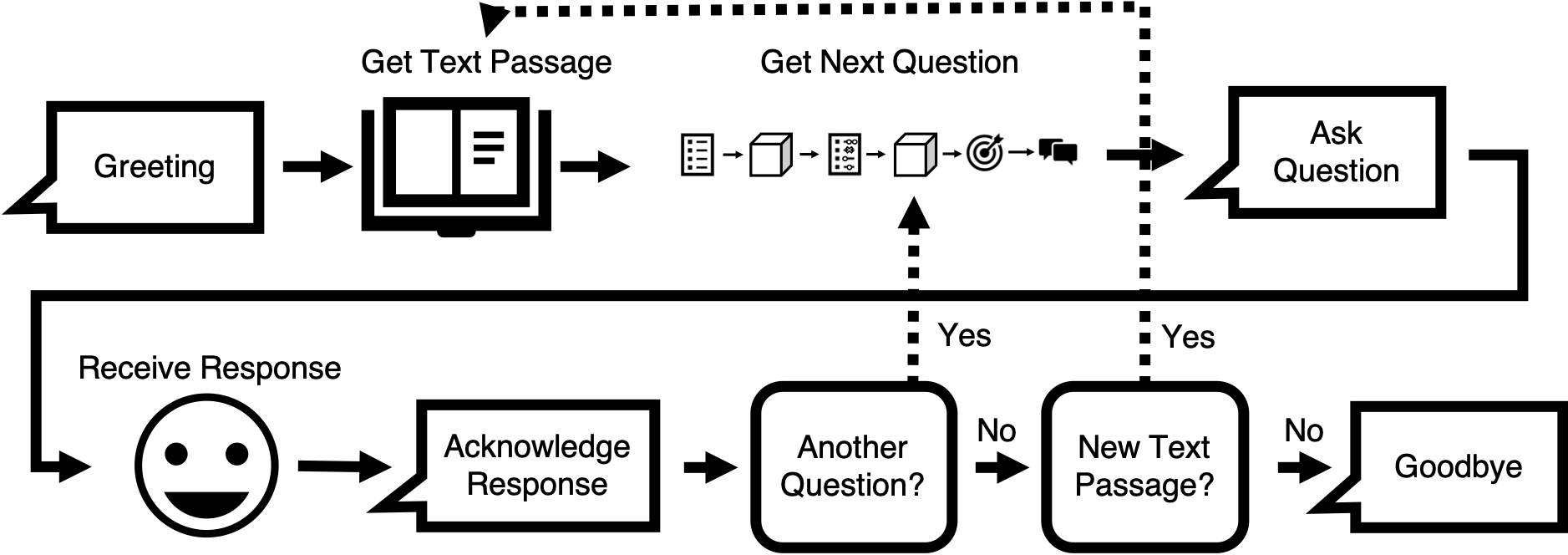}
	\caption{Dialogue system flow.}
	\label{dialogue_system}
\end{figure}

Our system converts raw text to questions about the \textit{novel metaphors}\footnote{Creative or unexpected metaphors, e.g., ``She \textit{frowned} like a \textit{thunderstorm}.''} within it using the pipeline in Figure \ref{processing_pipeline}.  It embeds the pipeline into the dialogue system in Figure \ref{dialogue_system}.  Our pipeline begins by segmenting text into sentences, and classifying each sentence as likely to contain a novel metaphor or not using a neural network model that considers sentence-level context\footnote{Words are represented using Word2Vec embeddings trained on Google News  \cite{Mikolov:2013:DRW:2999792.2999959}.} and psycholinguistic features \cite{parde-nielsen:aaai2018}.  We train the model using sentences from an existing dataset for which word pairs were labeled with metaphor novelty scores \cite{parde_lrec18}.  To repurpose the dataset for sentence-level classification, we label each sentence with a binary value depending on whether any word pairs within it exceed a threshold novelty score.

The system extracts all syntactically-related pairs of \textit{content words} (nouns, verbs, adjectives, and adverbs) from sentences classified as likely to contain a novel metaphor, and predicts a metaphor novelty score for each word pair using our scoring approach defined in prior work \cite{parde-nielsen:aaai2018}.  We train our scoring model on a combination of datasets: (1) the only publicly available metaphor novelty dataset, consisting of continuous metaphor novelty scores for 18,439 word pairs from multiple domains \cite{parde_lrec18}, and (2) a smaller dataset of 2100 word pairs extracted from Project Gutenberg (\url{www.gutenberg.org}) books, for which we crowdsourced ratings along the same continuous scale.  

Finally, the system selects a word pair based on chronological order, predicted novelty score, similarity to word pairs for which questions have already been generated, similarity of the word pair's source sentence to those for which questions have already been generated, and estimated completion time.  It generates a question for the selected word pair using the template-based \textit{Questioning the Author} (QtA) framework \cite{parde2018inlg}.  QtA is a questioning technique that prompts readers to consider the author's underlying motivations in crafting prose \cite{beck2006improving}.  We previously showed that automatically-generated QtA questions are cognitively deep and comparable to those generated by humans about the same topics \cite{parde2018inlg}.

\section{Usability Evaluation}
Twenty-six participants each discussed Jane Austen's \textit{Pride and Prejudice}\footnote{\textit{Pride and Prejudice} is the most-downloaded book on Project Gutenberg.} with our learning companion robot for 1-3 separate, approximately 30-minute sessions.
The system was implemented on a NAO robot named Grace, and we encoded contextual gestures and life-like swaying motions to facilitate natural interactions.  We employed a Wizard-of-Oz speech recognition technique, but all other aspects of the system functioned autonomously.  Following interaction sessions, participants were asked to rate their agreement on a five-point Likert scale ranging from 1 (Strongly Disagree) to 5 (Strongly Agree) for each of nine statements (provided in Table \ref{survey_results}) covering different aspects of the interaction.



Twenty-six (12 M/14 F) participants completed one interaction session, 18 (9 M/9 F) additionally completed a second, and 7 (4 M/3 F) completed a third.  The survey results, including the mode and median (ties broken by averaging) scores, the 95\% confidence interval for each survey statement, and the $p$ values resulting from a one sample $t$-test that compared the sample mean to an expected population mean of 3.0 (``Neither Agree Nor Disagree'') are shown in Table \ref{survey_results}. 
All average scores expressed positive sentiment, and most differences between the average and the Likert scale midpoint (3.0) were statistically significant.  The results establish that adults are receptive to an embodied lifelong learning companion, persistently rating it as both likeable and engaging.  No sharp reductions in scores were observed over repeated sessions, which may suggest that the interactions are engaging enough to appeal to users for regular use.

\begin{table}[t]
	\scriptsize
	\centering
	\caption{Mode, median, 95\% confidence interval, and $p$ for each statement, for each session.  For sessions 1, 2, and 3, $n$=26, $n$=18, and $n$=7, respectively.}
	\begin{tabular}{l|c|c|c||c|c|c||C{1cm}|C{1cm}|C{1cm}||c|c|c|}
		\cline{2-13}
		& \multicolumn{3}{c||}{\textbf{Mode}}   & \multicolumn{3}{c||}{\textbf{Median}} & \multicolumn{3}{c||}{\textbf{95\% C.I.}} & \multicolumn{3}{c|}{\textit{\textbf{p}}} \\ \hline
		\multicolumn{1}{|l|}{\textbf{Statement}}                                                       & \textbf{1} & \textbf{2} & \textbf{3} & \textbf{1} & \textbf{2} & \textbf{3} & \textbf{1}  & \textbf{2} & \textbf{3}   & \textbf{1}   & \textbf{2}  & \textbf{3}  \\ \hline
		\multicolumn{1}{|L{4.7cm}|}{S1: I found Grace easy to understand.}                                    & 4          & 4          & 5          & 4.0        & 4.0        & 4.0        & 3.9 $\pm$ .3  & 3.9 $\pm$ .5 & 3.9 $\pm$ .8   & .00          & .00         & .11         \\ \hline
		\multicolumn{1}{|L{4.7cm}|}{S2: I knew what I could say or do at each point of the dialogue.}         & 3          & 3          & 4          & 3.5        & 3.5        & 4.0        & 3.5 $\pm$ .4  & 3.6 $\pm$ .4 & 4.1 $\pm$ .5   & .04          & .02         & .00         \\ \hline
		\multicolumn{1}{|L{4.7cm}|}{S3: The system worked the way I expected.}                                & 4          & 3          & 5          & 4.0        & 3.5        & 5.0        & 3.7 $\pm$ .3  & 3.7 $\pm$ .5 & 4.6 $\pm$ .4   & .00          & .02         & .00         \\ \hline
		\multicolumn{1}{|L{4.7cm}|}{S4: I would like to use this system regularly.}                           & 3          & 4          & 3          & 4.0        & 3.5        & 3.0        & 3.6 $\pm$ .4  & 3.3 $\pm$ .5 & 3.4 $\pm$ .7   & .00          & .33         & .29         \\ \hline
		\multicolumn{1}{|L{4.7cm}|}{S5: I like interacting with Grace.}                                       & 5          & 4          & 5          & 4.5        & 4.0        & 4.0        & 4.3 $\pm$ .3  & 3.8 $\pm$ .5 & 3.9 $\pm$ .8   & .00          & .01         & .11         \\ \hline
		\multicolumn{1}{|L{4.7cm}|}{S6: Grace seems smart.}                                                   & 3          & 3          & 3          & 3.0        & 3.0        & 4.0        & 3.5 $\pm$ .4  & 3.2 $\pm$ .5 & 3.7 $\pm$ .8   & .01          & .39         & .14         \\ \hline
		\multicolumn{1}{|L{4.7cm}|}{S7: Grace's dialogue seems natural.}                                      & 2          & 4          & 4          & 3.0        & 4.0        & 4.0        & 3.3 $\pm$ .4  & 3.3 $\pm$ .5 & 3.9 $\pm$ .7   & .23          & .25         & .08         \\ \hline
		\multicolumn{1}{|L{4.7cm}|}{S8: Grace asked interesting questions about the text we were discussing.} & 3          & 3          & 4          & 4.0        & 3.5        & 4.0        & 3.5 $\pm$ .5  & 3.6 $\pm$ .5 & 4.3 $\pm$  .5 & .04          & .03         & .00         \\ \hline
		\multicolumn{1}{|L{4.7cm}|}{S9: It made sense for Grace to ask the questions we discussed.}          & 4          & 4          & 5          & 4.0        & 4.0        & 5.0        & 3.8 $\pm$ .3  & 3.7 $\pm$ .5 & 4.6 $\pm$ .4   & .00          & .01         & .00         \\ \hline
	\end{tabular}
	\label{survey_results}
\end{table}

\section{Conclusions}
In this work, we design and implement an embodied lifelong learning companion that engages users cognitively via human-robot book discussions.  We conduct a usability evaluation of our prototype, and find that users rate the learning companion as likeable and engaging across multiple sessions.  Future work will focus on personalization and conversation quality, driving the system closer to our goal of automatically facilitating the types of conversations one might encounter during a cognitively stimulating book discussion with a human companion.

\section*{Acknowledgements}
This material was based upon work supported by a National Science Foundation Graduate Research Fellowship under Grant 1144248, and the National Science Foundation under Grant 1262860. Any opinions, findings, and conclusions or recommendations expressed in this material are those of the author(s) and do not necessarily reflect the views of the National Science Foundation.

\bibliography{aied19parde}
\bibliographystyle{splncs04}

\end{document}